\documentclass[twocolumn]{aastex7}

\usepackage{url}

\usepackage{amsmath}
\usepackage{float}
\usepackage{multirow}
\usepackage{soul}
\usepackage{rotating}
\usepackage{hyperref}
\usepackage{wrapfig}
\usepackage{graphicx,caption}
\usepackage{siunitx}
\usepackage{gensymb}
\usepackage[table]{xcolor}

\newcommand{\bjdtdb}{\ensuremath{\rm {BJD_{TDB}}}}
\newcommand{\feh}{\ensuremath{\left[{\rm Fe}/{\rm H}\right]}}

\newcommand{\teff}{\ensuremath{T_{\rm eff}}}

\newcommand{\logg}{\ensuremath{\log g_*}}

\newcommand{\msun}{\ensuremath{\,M_\Sun}}
\newcommand{\rsun}{\ensuremath{\,R_\Sun}}
\newcommand{\lsun}{\ensuremath{\,L_\Sun}}

\newcommand{\mj}{\ensuremath{\,M_{\rm J}}}
\newcommand{\mearth}{\ensuremath{\,M_{\Earth}}}

\newcommand{\mplanet}{\ensuremath{\,M_{\rm P}}}
\newcommand{\rplanet}{\ensuremath{\,R_{\rm P}}}

\newcommand{\rj}{\ensuremath{\,R_{\rm J}}}

\newcommand{\fave}{\langle F \rangle}
\newcommand{\fluxcgs}{10$^9$ erg s$^{-1}$ cm$^{-2}$}

\newcommand{\tess}{{\it TESS}}

\newcommand{\kms}{\,km\,s$^{-1}$}

\newcommand{\mstar}{\ensuremath{M_{*}}}
\newcommand{\rstar}{\ensuremath{R_{*}}}
\newcommand{\ar}{\ensuremath{a/R_*}}
\newcommand{\vsini}{\ensuremath{v\sin{i_\star}}}
\newcommand{\exofasttwo}{{\tt EXOFASTv2}}

\newcommand{\degrees}{\ensuremath{^{\circ}}}
\providecommand{\bjdtdb}{\ensuremath{\rm {BJD_{TDB}}}}
\providecommand{\tjdtdb}{\ensuremath{\rm {TJD_{TDB}}}}
\providecommand{\feh}{\ensuremath{\left[{\rm Fe}/{\rm H}\right]}}
\providecommand{\teff}{\ensuremath{T_{\rm eff}}}

\providecommand{\msun}{\ensuremath{\,M_\Sun}}
\providecommand{\rsun}{\ensuremath{\,R_\Sun}}
\providecommand{\lsun}{\ensuremath{\,L_\Sun}}
\providecommand{\mj}{\ensuremath{\,M_{\rm J}}}
\providecommand{\rj}{\ensuremath{\,R_{\rm J}}}

\providecommand{\fave}{\langle F \rangle}
\providecommand{\fluxcgs}{10$^9$ erg s$^{-1}$ cm$^{-2}$}
\providecommand{\finallam}{$-68\fdg1^{+7.5}_{-5.3}$}
\providecommand{\finalpsi}{$72\fdg2^{+6.4}_{-6.6}$}

\begin{document}

\title{Early Evidence for Polar Orbits of Sub-Saturns Around Hot Stars}

\author[0009-0007-3267-9088]{Emma Dugan}
\affiliation{Department of Astronomy, Indiana University, 727 East 3rd Street, Bloomington, IN 47405-7105, USA}
\email{dugane@iu.edu}

\author[0000-0002-0376-6365]{Xian-Yu Wang}
\altaffiliation{Sullivan Prize Postdoctoral Fellow}
\affiliation{Department of Astronomy, Indiana University, 727 East 3rd Street, Bloomington, IN 47405-7105, USA}
\email{xwa5@iu.edu}

\author[0009-0004-1149-9887]{Agustin Heron}
\affiliation{Department of Astronomy, Indiana University, 727 East 3rd Street, Bloomington, IN 47405-7105, USA}
\email{aguheron@iu.edu}

\author[0000-0002-5181-0463]{Hareesh Gautham Bhaskar}
\affiliation{Department of Astronomy, Indiana University, 727 East 3rd Street, Bloomington, IN 47405-7105, USA}
\email{hbhaskar@iu.edu}

\author[0000-0002-7670-670X]{Malena Rice}
\affiliation{Department of Astronomy, Yale University, 219 Prospect Street, New Haven, CT 06511, USA}
\email{malena.rice@yale.edu}

\author[0000-0003-0412-9314]{Cristobal Petrovich}
\affiliation{Department of Astronomy, Indiana University, 727 East 3rd Street, Bloomington, IN 47405-7105, USA}
\email{cpetrovi@iu.edu} 

\author[0000-0002-7846-6981]{Songhu Wang}
\affiliation{Department of Astronomy, Indiana University, 727 East 3rd Street, Bloomington, IN 47405-7105, USA}
\email{sw121@iu.edu}

 \correspondingauthor{Xian-Yu Wang}
 \email{xwa5@iu.edu}

\begin{abstract}
Sub-Saturns have been reported to preferentially occupy near-polar orbits, but this conclusion has so far been based primarily on systems with cool host stars; obliquity measurements for sub-Saturns orbiting hot stars remain scarce. Expanding the census into the hot-star regime is essential to test whether the polar preference persists across the Kraft break and to diagnose the underlying excitation mechanisms. In this work, we present Rossiter–McLaughlin observations of TOI-1135\,b, a sub-Saturn orbiting a hot star with \teff\,=\,6320 $\pm$ 120 K, using WIYN/NEID. We confirm its near-polar architecture, measuring a sky-projected obliquity of $\lambda=$\finallam, and a true obliquity of $\psi=$\finalpsi. Coupling our new measurement with stellar-obliquity data from the literature, we find that sub-Saturns and hot Jupiters around cool stars are unlikely to be drawn from the same parent distribution at the ${5.2}\sigma$ level, consistent with weaker tidal realignment induced by lower-mass planets. Of the two known misaligned sub-Saturns around hot stars, both are near-polar, suggesting that the polar preference may extend above the Kraft break. Moreover, their obliquities lie near $\sim65\degrees$, supporting predictions from secular resonance crossing for sub-Saturns around rapidly rotating hot stars.
\end{abstract}

\keywords{Exoplanet astronomy(486), exoplanet dynamics (490), exoplanet systems (484), exoplanets (498), planetary alignment (1243), planetary theory (1258), star-planet interactions (2177)}

\section{Introduction} \label{sec:intro}

Exoplanets undergo diverse dynamical processes during their formation and evolution, imprinting distinct architectural signatures on planetary systems. Among the key parameters that characterize such systems, stellar obliquity (the angle between the stellar spin axis and the planetary orbital axis, $\psi$) serves as a critical diagnostic tool. The distribution of stellar obliquity (see \citealt{Queloz2000}, \citealt{Winn2010}, \citealt{Schlaufman2010}, \citealt{Albrecht2012}, \citealt{Albrecht2022}, \citealt{Knudstrup2024}, and references therein) 
constrains both the \textit{origins} of misalignment --- via mechanisms capable of delivering hot Jupiters, such as planet–planet scattering \citep{Rasio1996, Beauge2012}, Kozai–Lidov oscillations \citep{Wu2003, Fabrycky2007, Naoz2016}, or other secular mechanisms like Coplanar High Eccentricity Migration  (CHEM, \citealt{Petrovich20152015CHEM}) and secular chaos \citep{Lee2003, WuLithwick2011, Li2014, LithwickWu2014, hamers_secular_2017, Teyssandier2019, Volk2020} --- as well as universal mechanisms that operate independently of close-in migration and across diverse system architectures, including chaotic accretion \citep{Bate2010, Thies2011, Fielding2015, Bate2018}, magnetic warping \citep{Foucart2011, Lai2011, Romanova2013, Romanova2021}, tilting by a companion star \citep{Borderies1984, Lubow2000, Batygin2012, Matsakos2017}, and internal-gravity-wave-induced tumbling \citep{Lai2012, Lin2017, Damiani2018}, consistent with disk-era observations suggesting that misalignment is already present in roughly one third of systems \citep{Biddle2025}. Beyond their origin, stellar obliquities can \textit{evolve} through tidal dissipation, often driving realignment through damping with strong dependencies on stellar structure and orbital separation \citep{Winn2010, Albrecht2012, Lai2012, Li2016, Wang2021, zanazzi2024damping,Zanazzi2025}.

Population-level studies reveal an overabundance of perpendicular systems (e.g., \citealt{Albrecht2021, Bourrier2022, Stefansson2022, Dong2023StellarObliquity, Siegel2023}) --- especially Jupiters ($0.3<\mplanet/\mj<13$) around F stars and sub-Saturns ($10<\mplanet/\mearth\le100$) \citep{Espinoza2024, Knudstrup2024}. Several competing explanations have been advanced, most notably: tidal damping, in some cases, can cause the obliquity to linger near 90\degrees \citep{Lai2012, RogersANDLin2013, Anderson2021}; von Zeipel Kozai Lidov (vZKL) cycles predict stellar obliquity peaks near 35\degrees\ and 115\degrees\ \citep{Fabrycky2007}; secular resonance crossing can yield polar orbits, especially for low-mass planets around cool stars \citep{Petrovich2020}; and magnetic warping can also produce polar configurations \citep{Foucart2011, Lai2011, Romanova2021}. Recently, more evidence has emerged linking distant companions, non-zero eccentricities, and radius inflation of sub-Saturns: companions have been detected in several misaligned systems (HAT-P-11, \citealt{Yee2018, Yee2024RNAAS, An2025}; WASP-107, \citealt{Piaulet2021}; HAT-P-7, \citealt{Yang2025}), potentially driving misalignment and sustaining eccentricity \citep{Yu2024, Lu2025}, thereby tidally heating and inflating misaligned sub-Saturns \citep{Sethi2025}; see also \citealt{Batygin2025} for an ohmic-heating alternative.

Despite recent progress, our understanding remains limited owing to the scarcity of measurements for sub-Saturns with hot stars above the Kraft break (\teff\,$\approx$ 6100 K, \citealt{kraft1967break}), which can be attributed to three main factors. First, sub-Saturns generally have smaller planetary radii than Jupiters, while hot stars tend to have larger stellar radii. The resulting small planet-to-stellar radius ratio (\rplanet/\rstar) leads to shallow transits, reducing the signal-to-noise ratio and making their discovery more challenging. Secondly, a small \rplanet/\rstar\, produces a weaker Rossiter–McLaughlin (RM) effect \citep{Rossiter1924, McLaughlin1924}, the most widely used method to measure stellar obliquity, further hindering $\psi$ measurement. Finally, the radial velocity precision for hot stars is generally lower than for cool stars, due to their fewer and broader spectral lines and higher levels of stellar jitter, making it even more difficult to measure the RM signal with sufficient accuracy.

Recent space missions, such as the Transiting Exoplanet Survey Satellite (\tess; \citealt{Ricker2015}), have discovered numerous sub-Saturn planets orbiting bright stars, making them well-suited for precise radial velocity measurements. Among them, TOI-1135~b \citep{mallorquin2024} stands out as an ideal system for measuring stellar obliquity. TOI-1135 b is a sub-Saturn with an orbital period of 8.03 days and a planetary mass of $\sim0.08\,M_{\rm J}$, orbiting a G0 bright star ($\teff\, = 6320\pm120$ K, V = 9.57 mag) that exhibits relatively fast stellar rotation ($P_{\rm rot} = 4.43 \pm 0.43$ days, $\vsini\, = 10.9^{+3.0}_{-2.1}$\kms).

In this work, we measured the RM effect of TOI-1135 b using the NEID spectrograph and found that its projected stellar obliquity is \finallam\, with a true stellar obliquity of \finalpsi. Following TOI-1842 (\teff\,=\,$6230\pm50$ K, $\lambda=-68{\degrees}^{+21}_{-15}$ , $\psi=73{\degrees}^{+16}_{-13}$, \citealt{Hixenbaugh2023}), it became the second misaligned, near-polar sub-Saturn orbiting a hot star. Meanwhile, this is also the 16$^{\rm th}$ result from the Stellar Obliquities in Long-period Exoplanet Systems (SOLES) survey \citep{Rice2021K2140, WangX2022WASP148, Rice2022WJs_Aligned, Rice2023Q6, Hixenbaugh2023, Dong2023, Wright2023, Rice2023TOI2202, Lubin2023, Hu2024PFS, Radzom2024, Ferreira2024, WangX2024, Radzom2025, Rusznak2025}

To place this stellar obliquity measurement in a broader context, we conducted a statistical analysis using the most recent stellar obliquity data. We quantified the difference between sub-Saturns and hot Jupiters around cool stars, finding that they are not drawn from the same population at the $3.6\sigma$ level. To date, there are three sub-Saturns around hot stars with stellar obliquity measurements; notably, two misaligned ones (TOI-1135 b and TOI-1842 b) are both near-polar. Although the sample size remains small, we speculate that the perpendicular pattern observed for sub-Saturns may persist even around hot stars. Finally, we find that the stellar obliquities of these two sub-Saturns are slightly offset from exact polar configurations, consistent with predictions from secular resonance crossing \citep{Petrovich2020}.

This paper is organized as follows. In Section~\ref{sec:obs}, we describe the observations of the system. Section~\ref{sec:SystemParameters} outlines the derivation of system parameters, including the determination of atmospheric stellar parameters using the NEID spectrum ($\S$~\ref{sec:iSpec}), the derivation of system parameters through global modeling ($\S$~\ref{sec:globalmodeling}), and the calculation of the true stellar obliquity ($\S$~\ref{sec:psi}). A statistical analysis is presented in Section~\ref{sec:Stats}, and we summarized our results in Section~\ref{sec:Summaryandimplication}.

\begin{figure*}
    \includegraphics[width=1\linewidth]{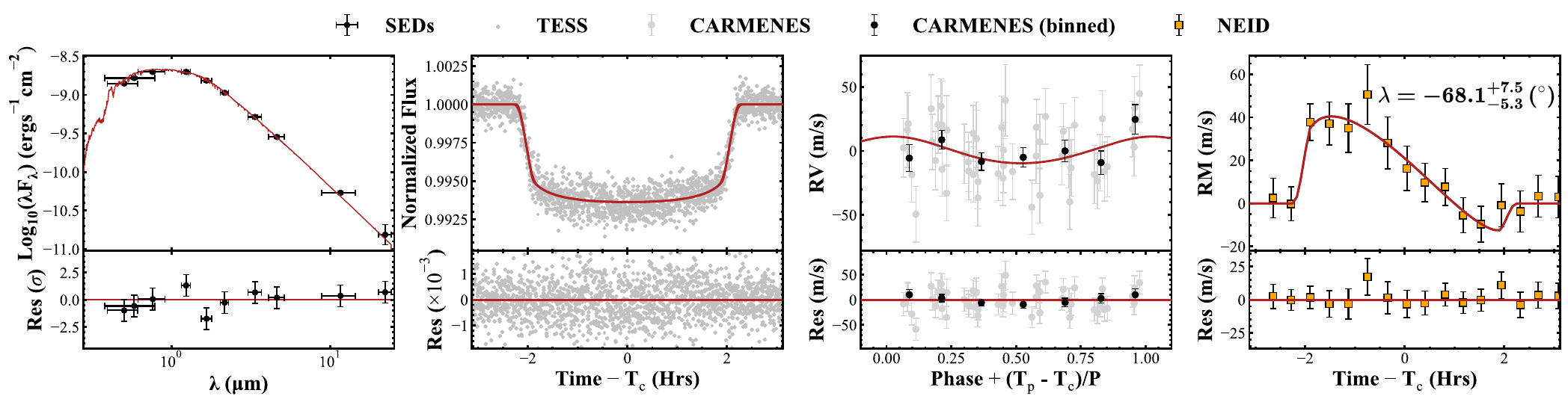}
    \caption{ Spectral energy distribution, \tess\ transit photometry, out-of-transit radial velocities, and Rossiter–McLaughlin effect modeling for TOI-1135. The upper panel in each plot shows the observational datasets along with the corresponding best-fit model from the EXOFASTv2 global modeling (solid red line). The resulting RM signal-to-noise ratio \citep{Kipping2024} is 8.1. The lower panel in each plot displays the residuals. (The data used to create this figure are \href{https://raw.githubusercontent.com/wangxianyu7/Data_and_code/refs/heads/main/TOI-1135RM/TOI1135.csv}{available}.)
    } 
    \label{fig:global_model}
\end{figure*}

\section{Observations} \label{sec:obs}

\subsection{\tess\, Photometry}
TOI-1135 was observed by the \tess\ satellite \citep{Ricker2015} in a total of 12 sectors spanning from 2019 to 2024, beginning with Sector 14 and continuing through Sector 86. These observations comprise 37 transits, including three partial transits at 2457390.65, 2463131.75, and 246642.95 (\bjdtdb). We adopted the TESS-Gaia Light Curve (TGLC\footnote{\url{https://github.com/TeHanHunter/TESS_Gaia_Light_Curve}}; \citealt{han2023}), which uses the position and magnitude of stars from Gaia DR3 \citep{GaiaCollaboration2023} to remove contamination from nearby stars and avoid potential underestimation of the transit depth, and thus the planetary radius \citep{Han2025}. The data used in this work include 200\,s cadence from Sectors 59, 74, and 79; 600\,s cadence from Sectors 40, 47, and 53; and 1800\,s cadence from Sectors 14, 19, 20, 21, and 26. The light curves from Sectors 14, 19, 20, 26, 40, and 47 were downloaded using the \texttt{lightkurve} package \citep{Lightkurve2018}, while the remaining light curves were obtained directly through the standard TGLC data extraction process, as they were not yet processed or uploaded to the Barbara A. Mikulski Archive for Space Telescopes (MAST).

\subsection{NEID Observation}

Spectroscopic observations of TOI-1135 were carried out on 2023 March 5 using the NEID spectrograph \citep{Schwab2016, Halverson2016} in its high-resolution mode ($R \sim 110{,}000$) on the 3.5\,m WIYN Telescope at Kitt Peak National Observatory. NEID is a fiber-fed \citep{Kanodia2018, Kanodia2023}, ultra-stable \citep{Stefannson2016, Robertson2019} spectrograph that has been operational since 2022 and covers a wavelength range of 380–930\,nm.

A total of 17 spectra were obtained between 02:48 and 08:32 UTC, using 1350\,s exposures. During the observations, the airmass ranged from $z = 1.66$ to $1.99$, and the seeing conditions varied between 1\farcs4\ and 3\farcs5, with a median of 1\farcs7.

The data were reduced using version 1.4.0 of the NEID Data Reduction Pipeline (\texttt{NEID-DRP})\footnote{Detailed documentation is available at \url{https://neid.ipac.caltech.edu/docs/NEID-DRP/}}. The resulting spectra achieved a median signal-to-noise ratio of 50 at 550\,nm, with a corresponding median radial velocity uncertainty of 10\,m\,s$^{-1}$. {The NEID RV data can be accessed through the link provided in the caption of Figure~\ref{fig:global_model}, and the corresponding spectra are available on the NEID Archive\footnote{\url{https://neid.ipac.caltech.edu/search.php}}.}

\section{System parameters}\label{sec:SystemParameters}
\subsection{Synthetic spectral fitting by \texttt{iSpec}}\label{sec:iSpec}

To derive the spectroscopic properties of the host star, including stellar effective temperature (\teff), surface gravity (\logg), metallicity (\feh), and projected rotational velocity (\vsini), we analyzed the co-added out-of-transit NEID spectra (S/N = 168) using the \texttt{iSpec} framework \citep{Blanco2014,Blanco2019}. Synthetic spectral fitting was performed following the same procedure described in Section~4.1 of \citet{WangX2024}. The final results of this spectroscopic analysis are reported in Table~\ref{tab:results}.

\subsection{\texttt{EXOFASTv2} global modeling}\label{sec:globalmodeling}

To derive consistent stellar and planetary parameters, we adopted a modified version of \texttt{EXOFASTv2} \citep{Eastman2013, Eastman2019} that incorporates a comprehensive RM model as described by \citet{Hirano2011}. We performed a global fit combining broadband photometry, \tess\ transit light curves, literature radial velocity (RV) measurements, and RM observations.

\noindent \ul{SED+MIST modeling}

{To derive stellar parameters including stellar effective temperature (\teff), radius (\rstar), mass (\mstar), metallicity (\feh), and age, we utilized the MESA Isochrones \& Stellar Tracks (MIST) model \citep{Choi2016mist, Dotter2016mist} in combination with a spectral energy distribution (SED) fitting approach, where \exofasttwo\, interpolates the MIST isochrones to generate model SEDs that are compared to the observed photometry in a self-consistent manner.} Photometry was compiled from various catalogs, including 2MASS \citep{Cutri2003}, WISE \citep{Cutri2014AllWISE}, \tess\, \citep{Ricker2015}, and Gaia DR3 \citep{GaiaCollaboration2023}. Gaussian priors based on our synthetic spectral fitting were applied to \feh, along with the parallax from Gaia DR3 and an upper limit for the $V$-band extinction from \citep{Schlafly2011}\footnote{\url{https://irsa.ipac.caltech.edu/applications/DUST/}}. A 2.4\% systematic uncertainty floor in \teff\, was adopted, as suggested by \cite{Tayar2022}.

\noindent \ul{Transit+RV modeling}

The 37 \tess\ transits, spanning 1974 days, enable a precise ephemeris and high-quality transit profile, which are critical for accurately determining the projected spin–orbit angle. Given the strong stellar activity reported by \citet{mallorquin2024}, the light curves were detrended using \texttt{wotan} \citep{Hippke2019}, employing the biweight method with an optimal window length of 12 hours (corresponding to 3$\times T_{14}$). The radial velocity measurements from CARMENES \citep{Quirrenbach2014,Quirrenbach2018} were taken from the discovery paper of TOI-1135b \citep{mallorquin2024}, while the NEID RM observations were obtained in this work.

To enable efficient exploration of the parameter space by the MCMC sampler, we adopted Gaussian priors only on \vsini\,, based on the results from the \texttt{iSpec} synthetic spectral fitting. Uniform priors were applied to the remaining parameters. To reduce the Lucy–Sweeney bias \citep{Lucy1971}, we parameterized the orbital eccentricity and argument of periastron using the combinations $\sqrt{e}\cos\omega$ and $\sqrt{e}\sin\omega$. For the same reason, we adopted the reparameterization $\sqrt{\vsini}\cos\lambda$ and $\sqrt{\vsini}\sin\lambda$ for the projected spin–orbit angle. Limb darkening coefficients were computed using stellar atmosphere models, and the V-band coefficients were applied to the NEID RM data. {In addition, we included RV jitter terms for the CARMENES and NEID data to account for RV variations caused by stellar activity.}

\noindent \ul{Parameters posterior derivation}

Posterior sampling was performed using the Parallel Tempering Differential Evolution Markov Chain Monte Carlo (PT-DE-MCMC) algorithm \citep{ter2006markov}. A total of 94 walkers were used, corresponding to twice the number of free parameters, along with eight temperature ladders. The MCMC process was considered converged when the Gelman–Rubin diagnostic \citep[$\hat{R}$;][]{Gelman1992} fell below 1.01 and the number of independent samples exceeded 1000. The final system parameters are summarized in Table~\ref{tab:results}.

\noindent \ul{Parameters comparison}

Compared with the parameters reported in the discovery paper \citep{mallorquin2024}, we find that all parameters are consistent with ours within 2$\sigma$, except for \logg\,(6$\sigma$) and \rplanet/\rstar\, (2.9$\sigma$). The discrepancy in $\logg$ may stem from an underestimated spectroscopic uncertainty in \citet{mallorquin2024}, as suggested by their unrealistically small \teff\, error (6150 $\pm$ 15 K; see also \citealt{Blanco2019} for a discussion of systematic biases in $\logg$ estimates across different spectroscopic analysis pipelines). Moreover, our \rplanet/\rstar\, value agrees well with that reported by \citet{Han2025} that adopted TGLC (0.5$\sigma$), suggesting that the use of TGLC data mitigates the contamination from nearby stars in the \tess\, light curve, enabling a more precise determination of $R_p/R_\star$ and thus the planetary radius. {In addition, the constraint on the stellar age from the \texttt{EXOFASTv2} SED+MIST fit is weak. Therefore, we adopt the stellar age of 125–1000 Myr from the discovery paper \citep{mallorquin2024}, which was derived using gyrochronology, NUV excess, kinematics, and lithium equivalent width, as the preferred estimate. Finally, the derived planetary mass of $0.083^{+0.051}_{-0.052}\,\mj$ is consistent with zero within $2\sigma$; therefore, we adopt $\mplanet < 0.185\,\mj$ as the $3\sigma$ upper limit and preferred value.}

The updated planetary radius, together with the planetary mass, indicates that TOI-1135\,b is a puffy exoplanet ($\rho_P = 0.15\pm0.09\,\mathrm{g\,cm^{-3}}$) with high incident flux ($F=310^{+28}_{-27}\,S_{\oplus}$; $T_{\mathrm{eq}}= 1171\pm25\,\mathrm{K}$), making it a promising target for future James Webb Space Telescope (JWST, \citealt{Gardner2006}) transmission spectroscopy and for further investigation of the relationship between sub-Saturn puffiness and misalignment.

\startlongtable
\begin{deluxetable*}{lccccccc}
\tablecaption{Median values and 68\% confidence interval for TOI-1135.}
\tabletypesize{\scriptsize}
\tablehead{\colhead{~~~Parameter} & \colhead{Description} & \multicolumn{6}{c}{Values}}
\startdata
\multicolumn{2}{l}{\textbf{Stellar parameters from iSpec fit:}}&\smallskip\\
~~~~$T_{\rm eff}$\dotfill &Effective temperature (K)\dotfill &$6150\pm102$\\
~~~~$[{\rm Fe/H}]$\dotfill &Metallicity (dex)\dotfill &$-0.06\pm0.07$\\
~~~~$\log{g}$\dotfill &Surface gravity (cgs)\dotfill &$4.48\pm0.16$\\
~~~~$\vsini$\dotfill & Projected rotational velocity (km s$^{-1}$)\dotfill &9.7$\pm$4.3\\
\hline
\multicolumn{2}{l}{\textbf{EXOFASTv2 fit (adopted):}}&\\
\multicolumn{2}{l}{Rossiter-McLaughlin Parameters:}&\smallskip\\
~~~~$V_{\zeta}$\dotfill &Macroturbulence dispersion (km s$^{-1}$)\dotfill &$5.2^{+1.2}_{-1.1}$\\
~~~~$V_{\xi}$\dotfill &Microturbulence dispersion (km s$^{-1}$)\dotfill &$1.5\pm0.9$\\
~~~~$\lambda$\dotfill &Projected Spin-orbit angle (Degrees)\dotfill &$-68.1^{+7.5}_{-5.3}$\\
~~~~$\vsini$\dotfill &Projected rotational velocity (km s$^{-1}$)\dotfill &$10.9^{+3.0}_{-2.1}$\\
\multicolumn{2}{l}{Stellar Parameters:}&\smallskip\\
~~~~$M_*$\dotfill &Mass (\msun)\dotfill &$1.119\pm0.069$\\
~~~~$R_*$\dotfill &Radius (\rsun)\dotfill &$1.202\pm0.037$\\
~~~~$L_*$\dotfill &Luminosity (\lsun)\dotfill &$2.08^{+0.22}_{-0.20}$\\
~~~~$\rho_*$\dotfill &Density (cgs)\dotfill &$0.908^{+0.084}_{-0.079}$\\
~~~~$\log{g}$\dotfill &Surface gravity (cgs)\dotfill &$4.328^{+0.028}_{-0.031}$\\
~~~~$T_{\rm eff}$\dotfill &Effective temperature (K)\dotfill &$6320\pm120$\\
~~~~$[{\rm Fe/H}]$\dotfill &Metallicity (dex)\dotfill &$-0.039^{+0.060}_{-0.062}$\\
~~~~$Age^{*}$\dotfill &Age (Gyr)\dotfill &$3.3^{+2.2}_{-1.6}$ {([0.125, 1])}\\
~~~~$EEP$\dotfill &Equal Evolutionary Phase \dotfill &$364^{+36}_{-26}$\\
~~~~$A_V$\dotfill &V-band extinction (mag)\dotfill &$0.29^{+0.12}_{-0.13}$\\
~~~~$\varpi$\dotfill &Parallax (mas)\dotfill &$8.774\pm0.015$\\
~~~~$d$\dotfill &Distance (pc)\dotfill &$113.98\pm0.19$\\
\vspace{-0.5cm}
\smallskip\\\multicolumn{2}{l}{Planetary Parameters:}&b\smallskip\\
~~~~$P$\dotfill &Period (days)\dotfill &$8.0277327^{+0.0000019}_{-0.0000018}$\\
~~~~$R_P$\dotfill &Radius (\rj)\dotfill &$0.885\pm0.028$\\
~~~~$M_P^{*}$\dotfill &Mass (\mj)\dotfill &$0.083^{+0.051}_{-0.052}$ {($<$0.185)}\\
~~~~$T_C$\dotfill &Time of conjunction (\tjdtdb)\dotfill &$2460466.30793\pm0.00018$\\
~~~~$a$\dotfill &Semi-major axis (AU)\dotfill &$0.0815^{+0.0016}_{-0.0017}$\\
~~~~$i$\dotfill &Inclination (Degrees)\dotfill &$89.30^{+0.19}_{-0.24}$\\
~~~~$e$\dotfill &Eccentricity \dotfill &$0.062^{+0.12}_{-0.045}$\\
~~~~$\omega_*$\dotfill &Arg of periastron (Degrees)\dotfill &$-10^{+100}_{-150}$\\
~~~~$T_{\rm eq}$\dotfill &Equilibrium temp (K)\dotfill &$1171\pm25$\\
~~~~$K$\dotfill &RV semi-amplitude (m/s)\dotfill &$8.0^{+4.7}_{-4.9}$\\
~~~~$R_P/R_*$\dotfill &Radius of planet in stellar radii \dotfill &$0.07566^{+0.00022}_{-0.00020}$\\
~~~~$a/R_*$\dotfill &Semi-major axis in stellar radii \dotfill &$14.57\pm0.44$\\
~~~~$T_{14}$\dotfill &Total transit duration (days)\dotfill &$0.18590^{+0.00043}_{-0.00039}$\\
~~~~$T_{FWHM}$\dotfill &FWHM transit duration (days)\dotfill &$0.17236^{+0.00034}_{-0.00033}$\\
~~~~$b$\dotfill &Transit impact parameter \dotfill &$0.176^{+0.062}_{-0.046}$\\
~~~~$\Theta$\dotfill &Safronov Number \dotfill &$0.0137^{+0.0082}_{-0.0085}$\\
~~~~$\fave$ in cgs\dotfill &Incident Flux (\fluxcgs)\dotfill &$0.421^{+0.038}_{-0.037}$\\
~~~~$\fave$ in S$_{\oplus}$\dotfill &Incident Flux (S$_{\oplus}$)\dotfill &$310^{+28}_{-27}$\\
~~~~$M_P/M_*$\dotfill &Mass ratio \dotfill &$7.1^{+4.3}_{-4.4} \times 10^{-5}$\\
\multicolumn{2}{l}{Wavelength Parameters:}&TESS&V (RM)\smallskip\\
~~~~$u_{1}$\dotfill &Linear limb-darkening coeff \dotfill &$0.185^{+0.018}_{-0.019}$&$0.364^{+0.051}_{-0.050}$\\
~~~~$u_{2}$\dotfill &Quadratic limb-darkening coeff \dotfill &$0.270\pm0.027$&$0.307^{+0.049}_{-0.051}$\\
\multicolumn{2}{l}{Telescope Parameters:}&CARMENES&NEID\smallskip\\
~~~~$\gamma_{\rm rel}$\dotfill &Relative RV Offset (m/s)\dotfill &$1.0^{+3.3}_{-3.2}$&$-21377.6^{+2.6}_{-2.5}$\\
~~~~$\sigma_J$\dotfill &RV Jitter (m/s)\dotfill &$20.8^{+2.9}_{-2.4}$&$0.00$\\
\enddata
\tablenotetext{}{\hspace{-1.5cm} $^{*}${Note: the values in parentheses are preferred, see discussion in Section 3.2.} }
\label{tab:results}
\end{deluxetable*}
\clearpage
 
\subsection{Stellar rotation period and true stellar obliquity}\label{sec:psi}

We performed Gaussian Process modeling of the transit-masked \tess\, light curves for TOI-1135 using the stellar rotation kernel described in Equation 56 of \citet{ForemanMackey2017} to derive the stellar rotation period and constrain the stellar inclination, which, in turn, allowed us to estimate the true 3D obliquity of the system. This analysis yielded a stellar rotation period ($P_{\mathrm{rot}}$) of $4.34 \pm 0.09$ days. Assuming a conservative 10\% uncertainty to account for spot evolution and differential rotation \citep{Epstein2014, Aigrain2015}, we adopted a final value of $P_{\mathrm{rot}} = 4.34 \pm 0.43$ days.

Using the \vsini\, and stellar radius derived from our global modeling, along with the stellar rotation period, we inferred the posterior distribution of the stellar inclination following the method described by \citet{Masuda2020stincl}. The resulting posterior shows $i_* = 59\fdg5^{+19.9}_{-18.1}$.

The true stellar obliquity, $\psi$, was calculated using the following relation \citep{Fabrycky2009}:
\begin{equation*} \label{eq:1}
\cos{\psi} = \cos{i_{*}}\cos{i} + \sin{i_{*}}\sin{i}\cos{\lambda}
\end{equation*} 
where $i_\star$ is the stellar inclination, $i$ is the orbital inclination, and $\lambda$ is the sky-projected spin–orbit angle. Combining the posteriors on stellar inclination, planetary inclination, and projected spin-orbit angle, we derived a true obliquity for TOI-1135 of $\psi =$\finalpsi, indicating a significantly misaligned near-polar system.

{To complement our RM + \vsini + $P_{\mathrm{rot}}$ analysis, we also performed a Reloaded Rossiter–McLaughlin (RRM; \citealt{Cegla2016ReloadedRM}) analysis to constrain the true stellar obliquity, as well as the center-to-limb convective variation (CLV) coefficients ($c_1$, $c_2$) and the differential-rotation factor ($\alpha_{\mathrm{rot}}$). Details of this analysis are provided in Appendix~\ref{appendix:RRM}. The preferred solution yields $\lambda = -74\fdg9^{+2.9}_{-2.7}$, $i\star = 49\fdg6^{+26.4}_{-34.1}$, $\psi = 78\fdg8^{+6.8}_{-4.2}$, $\alpha_{\mathrm{rot}} = 0.19^{+0.16}_{-0.13}$, $c_1 = -4.5^{+4.0}_{-3.1}$, and $c_2 = 0.8^{+2.1}_{-2.7}$. The $\lambda$, $i\star$, and $\psi$ values are consistent with those derived from the RM + \vsini + $P_{\mathrm{rot}}$ analysis within 1.2$\sigma$. Assuming the periodic photometric modulation originates from starspots at a latitude of $45\degree$, the expected rotation period is $P_{45\degree} = P_{\mathrm{eq}} / (1 - \alpha_{\rm rot} \sin^2 45\degree)$, which is 10.5\% longer than the equatorial rotation period ($P_{\mathrm{eq}}$), validating the adopted 10\% uncertainty for $P_{\mathrm{rot}}$.
}

\section{Statistical Analysis and Discussion}\label{sec:Stats}
Benefiting from the rapidly increasing number of detected sub-Saturns --- boosted by the \tess\ mission and the advent of extremely high-precision spectrographs (e.g., NEID; \citealt{schwab2016design}, ESPRESSO; \citealt{Pepe2021}, and KPF; \citealt{Gibson2024}) --- the population of sub-Saturns with stellar obliquity measurements has expanded swiftly, enabling a series of population-level analyses \cite[e.g.,][]{Stefansson2022,Hixenbaugh2023,Bourrier2023,Attia2023,Knudstrup2024,Espinoza2024}. However, sub-Saturns orbiting hot stars remain largely unexplored. In this work, we present the second misaligned, near-polar ($\lambda = $\finallam, $\psi = $\finalpsi) sub-Saturn around hot star, TOI-1135 b. With our new measurements, we conduct a population-level comparison of stellar obliquities between sub-Saturns and hot Jupiters to examine potential differences and emerging trends.

\subsection{Sample construction}\label{sample}

We cross-matched the TEPCat stellar obliquity table\footnote{\url{https://www.astro.keele.ac.uk/jkt/tepcat/obliquity.html}, {accessed on October 04, 2025}} \citep{Southworth2011} with composite planet data table \citep{NEA2} from the NASA Exoplanet Archive (NEA; \citealt{Christiansen2025}). In our catalog, \teff\, and $\lambda$ were adopted from TEPCat, while planetary masses were derived from the NEA. 

We excluded systems with low-quality or contested obliquity measurements (see Appendix B of \citealt{WangX2024}). Additionally, to minimize observational biases, we restricted our sample to stellar obliquity measurements derived using the RM effect, with priority given in the following order: classical RM, Doppler shadow \citep{Albrecht2007,Collier2010,Zhou2016,Johnson2017}, reloaded RM \citep{Cegla2016ReloadedRM}, and RM Revolution \citep{Bourrier2021RMrevolutions} methods. {Systems in compact multi-planet configurations ($P_{2}/P_1 <6$, see e.g., \citealt{WangX2022WASP148}) were also excluded,} as their dynamical histories may differ significantly from those of single-planet systems \citep{Albrecht2013, Wang2018a, Zhou2018, Dai2023, Radzom2024, Radzom2025}, potentially complicating the interpretation of stellar obliquities. Our sample was then pared down to include only hot Jupiters ($\ar <10$, $0.3 < \mplanet/\mj < 13$) and sub-Saturns ($10 < \mplanet/\mearth \leq 100$). These categories were then divided further into hot and cool stars by defining the divisor as the Kraft break (\teff\,$\approx$6100 K). The final sample includes {30} sub-Saturns around cool stars and 3 around hot stars\footnote{
{The planetary mass of TOI-1859 b has been refined, now exceeding 0.3 $M_{\mathrm J}$ at the 2 $\sigma$ level, based on new NEID RV measurements by our collaboration (in prep.)}},
and {47} hot Jupiters around cool stars and {88} around hot stars, respectively. For the statistical analysis, we adopt $\lambda$ as a proxy for $\psi$, a well-validated proxy in spin–orbit studies that captures the main dynamical trends (e.g., \citealt{Fabrycky2009, Winn2010, Albrecht2012, Albrecht2022, Knudstrup2024}).

\begin{figure*}
    \centering
    \includegraphics[width=\textwidth]{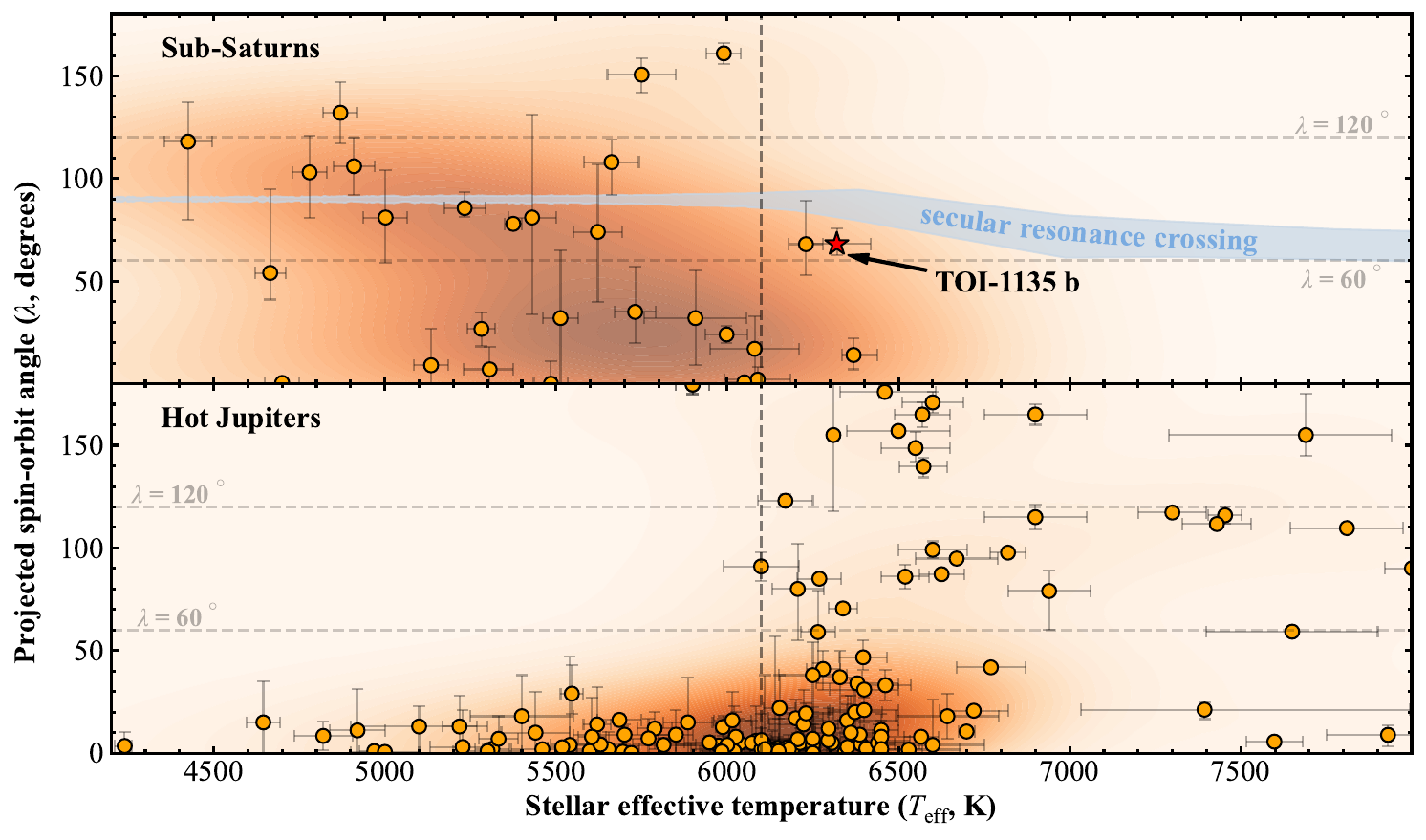}
    \caption{
        Projected spin–orbit angle as a function of stellar effective temperature for transiting exoplanets.
        The top panel shows sub-Saturns, while the bottom panel displays hot Jupiters.
        The vertical dashed line indicates the Kraft break at 6100 K, separating stars with convective envelopes (cooler) from those with radiative envelopes (hotter). TOI-1135 b is marked with a red star and labeled for reference. The predicted relationship between \teff\, and $\lambda$ from secular resonance crossing \citep{Petrovich2020} is shown as the shaded gray region.}
    \label{fig:teff_lam}
\end{figure*}

\subsection{Sub-Saturn can be misaligned with cool stars, while hot Jupiters are not}\label{subsection:temperature}

The most well-known relation concerning stellar obliquity is that hot Jupiters can be misaligned around hot stars but are typically aligned with their cool hosts \citep{Winn2010, Schlaufman2010, Albrecht2012, Knudstrup2024}. In contrast, sub-Saturns exhibit significant misalignment around cool stars. To quantify the difference between the cool-star stellar obliquity of sub-Saturn and hot Jupiters, we performed the two-sample Kolmogorov-Smirnov (K--S, \citealt{Hodges1958TheSP}) and Anderson-Darling tests (A--D, \citealt{scholz1987k}), implemented in \texttt{scipy} \citep{virtanen2020scipy}, on the stellar obliquity distributions for sub-Saturns and hot Jupiters. 

For the K–S and A–D analyses, we consider two cases: with and without incorporating measurement uncertainties. A direct comparison, without accounting for the uncertainties in $\lambda$, of the sky-projected obliquities of sub-Saturns and hot Jupiters around cool stars yields significance levels of {5.2}$\sigma$ and {4.4}$\sigma$ from the K–S and A–D tests, respectively. As a complementary analysis, we performed a Monte Carlo assessment that incorporates the measurement uncertainties in $\lambda$. For each system, we drew a single $\lambda$ value from a Gaussian distribution centered on the measured value, using the larger of the upper or lower uncertainties as the standard deviation to remain conservative. Using these simulated datasets, we conducted K--S and A--D tests to compare the $\lambda$ distributions of sub-Saturns and hot Jupiters around cool stars. This procedure was repeated 100,000 times to build distributions of p-values and significance levels, ensuring statistical robustness against measurement uncertainties. {We found that in 99\% of the iterations, both the K–S and A–D tests yielded significance levels greater than 3$\sigma$.} Therefore, whether or not uncertainties are taken into account, the significance levels remain sufficiently high to support a substantial difference in the stellar obliquity distributions of sub-Saturns and hot Jupiters around cool stars.

This difference supports the action of tidal realignment for hot Jupiters around cool stars: their high planetary masses, combined with the thick convective envelopes and strong resonance-locking effect of cool host stars, result in tidal dissipation strong enough to realign the stellar spin. The strength of this effect is reflected in the tidal realignment timescale, $\tau \propto (\mplanet/\mstar)^{-2}$ \citep{zahn1977reprint, Albrecht2012}. In contrast, the relatively small masses of sub-Saturns cannot drive efficient tidal dissipation, allowing them to remain misaligned even around cool stars.

\subsection{Sub-Saturns can be polar, even above Kraft break }\label{subsection: polar}

To date, stellar obliquity has been measured for only three sub-Saturn systems around hot stars. Two of them, TOI-1135 (this work) and TOI-1842 \citep{Hixenbaugh2023}, are significantly misaligned, exhibiting nearly polar orbits. This suggests that sub-Saturns can retain polar configurations even beyond the Kraft break. The two measured projected obliquities associated with polar orbits fall within a narrow range centered at $65\degrees$, with a combined 1$\sigma$ scatter of $15\degrees$, which is consistent with an exact 90\degrees\ polar orbit within 2$\sigma$. These measurements provide early statistical evidence for a concentrated population of polar orbits among sub-Saturns around hot stars.

The presence of high obliquity in sub-Saturns around hot stars offers valuable leverage for distinguishing between formation and migration pathways. {Although tidal damping driven by convective envelopes can, under certain conditions, can maintain obliquities near $90\degrees$ \citep{Lai2012, RogersANDLin2013, Anderson2021}, hot stars lack substantial convective envelopes and are therefore unlikely to produce or maintain polar orbits for sub-Saturns around such hosts.}

\subsection{Potential anti-correlation between \teff\,and $\lambda$ \\ in near-polar sub-Saturn systems}\label{subsection: unticorrolection}

Beyond identifying polar sub-Saturns around hot stars, we find that, in these systems, stellar obliquities might be systematically lower, suggesting a possible dependence on \teff. To quantify this trend, we performed a Pearson correlation analysis for {all} misaligned sub-Saturns with $\lambda > 50^\circ$ and found a correlation coefficient of $r = -0.6$ with a $p$-value of 0.04 ($< 0.05$). {Assuming a uniform distribution between $50\degrees$ and $180\degrees$, the probability that two independent systems would both fall within this 1$\sigma$ range ($50\degrees$–$80\degrees$) is $p=(30/130)^2=5.3\%$. }

{This result provides tentative support for an anti-correlation between \teff\ and $\lambda$. Both measurements show stellar obliquities slightly smaller than 90 degrees, which seem consistent with the prediction from secular resonance crossing. Additional RM measurements are required to confirm or refute this potential trend. Detecting four more sub-polar sub-Saturns around hot stars ($\lambda\sim65\degrees$ and \teff\,$\sim$ 6250 K) would reduce the $p$-value from the Pearson correlation analysis to below 0.01, corresponding to a $3\sigma$ significance.}

{Sub-polar ($60\degree \lesssim \psi$ [or $\lambda$] $< 90\degree$) sub-Saturns around hot stars would be consistent with predictions from secular resonance crossing \citep{Petrovich2020}, which occurs during the protoplanetary disk dispersal phase lasting $\sim$0.2–1 Myr, and is thus applicable for young systems such as TOI-1135.} When general relativistic apsidal precession dominates over the stellar quadrupole term, the stellar obliquity can be driven toward 90\degrees; in contrast, if the precession is dominated by the stellar quadrupole, the obliquity growth would be detuned at a sub-polar value, approaching a critical angle of $\cos^{-1} (\sqrt{1/5})\simeq 63\fdg4$ for very rapidly rotating stars (see our simulations in Figure~\ref{fig:teff_lam}  at high \teff\,). Hot stars typically maintain faster rotation and thus stronger stellar quadrupole, making them good laboratories for testing this mechanism. Although the current sample is limited, the two $\sim$65\degrees\ misaligned sub-Saturns around hot stars provide tentative evidence supporting this scenario. {Furthermore, secular resonance crossing predicts that any outer companion with a Jupiter mass and $a_{\rm out}\gtrsim0.75(a_{\rm in}/0.05)^{4/3}$ AU would be sufficient to set sub-Saturns beyond the Kraft break to sub-polar configurations. Additional RV observations and the upcoming Gaia DR4 epoch data will help confirm or rule out their existence.}

\section{Summary}\label{sec:Summaryandimplication}

The stellar obliquities of Jovian planets have been extensively measured, owing to their strong photometric and spectroscopic signals, yielding a comparatively large sample for investigating the origin and evolution of spin–orbit architectures. Hot Jupiters, which dominate this sample, exhibit a well-established \teff–$\lambda$ trend: systems around cool stars tend to be aligned, whereas those around hot stars often show significant misalignment \citep{Winn2010, Schlaufman2010, Albrecht2012, Knudstrup2024}. Their tidally detached analogs, warm Jupiters, in contrast, are generally aligned even around hot stars \citep{Rice2022WJs_Aligned, WangX2024}.

By comparison, the sample of sub-Saturns with measured obliquities remains small, especially for hot hosts, hindering our understanding of their underlying stellar obliquity distribution as well as the mechanisms driving misalignment excitation. To help fill this sparse regime, we present Rossiter–McLaughlin observations of the sub-Saturn TOI-1135\,b, a rare example orbiting a hot star, measuring a sky-projected obliquity of $\lambda=$\finallam\ and a true obliquity of $\psi=$\finalpsi.

Placing our stellar obliquity measurement of TOI-1135 in the context of the updated sample reveals three noteworthy patterns:
\begin{itemize}
    \item Compared to hot Jupiters around cool stars, sub-Saturns can display a higher rate of misalignment around cool stars at the 3.6$\sigma$ level, consistent with weaker tidal realignment induced by lower-mass planets.
    \item Among the three hot-star sub-Saturns with secure obliquities, two, including TOI-1135 b and TOI-1842 b, are near-polar, suggesting that the polar preference observed for cool hosts may persist beyond the Kraft break.
    \item  The two hot-star sub-Saturns do not sit exactly at 90\degrees\ but instead cluster near 65\degrees, consistent with predictions from secular resonance crossing.
\end{itemize}

{Further investigation of the mechanisms that trigger spin–orbit misalignment will require enlarging the sub-Saturn obliquity sample across the full stellar temperature range, with particular emphasis on hot hosts where tidal realignment is negligible and dynamical processes dominate. For instance, the sub-polar pattern occur preferentially in sub-Saturns, since triggering misalignment in more massive hot Jupiters via secular resonance crossing would require an outer companion that is both significantly inclined and massive, a configuration that is dynamically unlikely \citep{Becker2017}.}

{Furthermore, whether the obliquity distributions of sub-Saturns and hot Jupiters are intrinsically similar remains an open question. Recent work \citep[e.g.,][]{Knudstrup2024} has hinted at a clustering near $90\degrees$ among sub-Saturns and hot Jupiters orbiting F-type stars, although their sub-Saturn sample was limited to cool hosts. Our results extend this emerging trend to hot-star systems and reveal tentative evidence for a sub-polar ($60\degree \lesssim \psi$ [or $\lambda$] $< 90\degree$) tendency.}

{Expanding this sample will clarify whether the apparent clustering of near-polar geometries among hot sub-Saturns reflects a genuine, mass-dependent outcome of secular resonance crossing or merely small-number statistics.} Additional obliquity measurements, combined with constraints on outer perturbers, will be essential for identifying the dominant excitation pathway. Moreover, TOI-1135,b has an inflated radius relative to its mass and receives a high incident flux of about $310\,S_\oplus$, making it an attractive target for \textit{JWST} atmospheric characterization and for exploring possible links between radius inflation and orbital misalignment.

\begin{acknowledgments}
We are grateful to the anonymous reviewer for their constructive feedback, which has helped us improve this manuscript substantially.
Xian-Yu thanks Jie Yu for the insightful discussion on stellar structure. We acknowledge support from the NASA Exoplanets Research Program NNH23ZDA001N-XRP (Grant No. 80NSSC24K0153) and  NASA TESS General Investigator - Cycle 7 NNH23ZDA001N-TESS (Grant No. 80NSSC25K7912). Additionally, M.R. and S.W. acknowledge support from the Heising-Simons Foundation, with M.R. supported by Grant $\#$2023-4478, and S.W. supported by Grant $\#$2023-4050. M.R. acknowledges support from National Geographic grant $\#$EC-115062R-24. This research was supported in part by Lilly Endowment, Inc., through its support for the Indiana University Pervasive Technology Institute.

{The \tess\, data from Sectors 14, 19, 20, 26, 40, and 47 used in this paper are available from the \tess\, Gaia Light Curve High Level Science product at MAST \citep{10.17909/610m-9474}. The remaining \tess\, data were obtained via the TGLC pipeline.
}
\end{acknowledgments}

\vspace{5mm}
\facilities{WIYN/NEID, TESS, NASA Exoplanet Archive}

\software{
\texttt{EXOFASTv2} \citep{Eastman2017, Eastman2019}, 
\texttt{matplotlib} \citep{hunter2007matplotlib}, 
\texttt{numpy} \citep{oliphant2006guide, walt2011numpy, harris2020array}, 
\texttt{pandas} \citep{mckinney2010data}, 
\texttt{scipy} \citep{virtanen2020scipy}, 
\texttt{statsmodels} \citep{seabold2010statsmodels},
{\texttt{PyORBIT} \citep{Malavolta2016,Malavolta2018}}
}
\clearpage
\appendix
\begin{figure}
    \centering
    \includegraphics[width=0.46\linewidth]{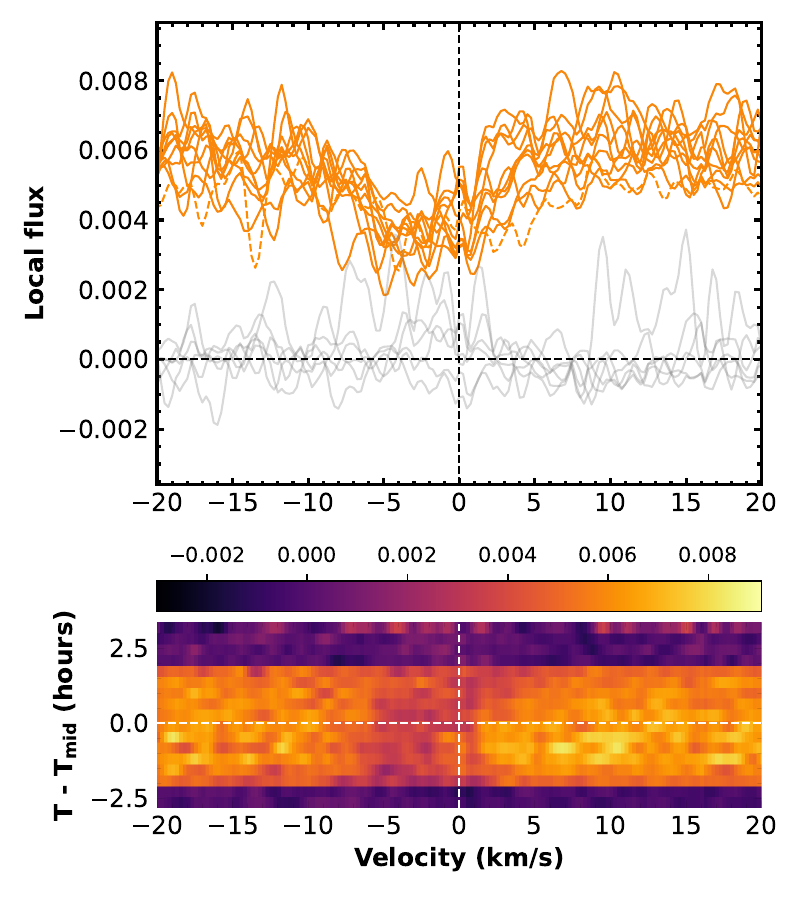}
    \includegraphics[width=0.49\linewidth]{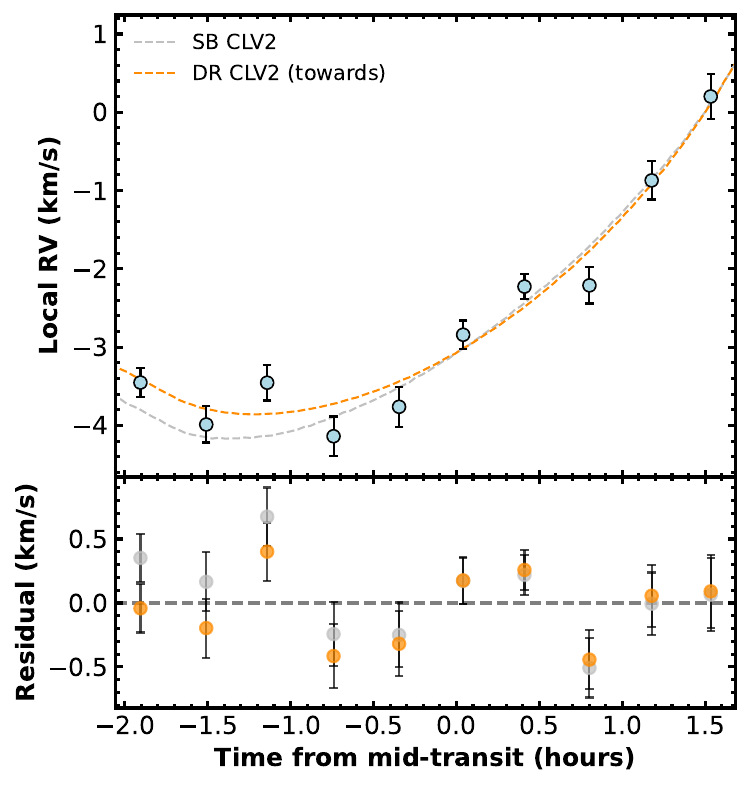}
\caption{{\textit{{Top left:}} Local CCFs (out-of-transit - in-transit) in the stellar rest frame behind TOI-1135\,b. The solid orange lines represent local CCFs at disk positions with $\langle \mu \rangle \geq 0.4$, dashed orange lines correspond to $0 < \langle \mu \rangle < 0.4$, and gray lines indicate out-of-transit regions. \textit{{Bottom left:}} A two-dimensional map of the local CCFs, color-coded by local flux. \textit{{Top right:}} Local RVs derived from local CCFs where $\langle \mu \rangle \geq 0.4$, with the two best-fit models (orange: differential rotation + centre-to-limb quadric model with $i_\star<90\degrees$; grey: solid body + centre-to-limb quadric model ) shown as dashed lines. \textit{{Bottom right:}} Residuals corresponding to the two models shown above.}}
    \label{fig:rrm}
\end{figure}
\section{{Reloaded Rossiter-McLaughlin effect analysis}}\label{appendix:RRM}

{The bright ($V = 9.57$) and rapidly rotating star TOI-1135 (\vsini\ = 10.9\,\kms), combined with its relatively large $\rplanet/\rstar$, makes it an excellent candidate for a Reloaded Rossiter–McLaughlin (RRM; \citealt{Cegla2016ReloadedRM}) analysis to constrain its true stellar obliquity, as well as to probe possible center-to-limb convective variations and differential rotation \citep{Roguet2022}. We followed the procedures described by \citet{Cegla2016ReloadedRM} and \citet{Doyle2025} to isolate the starlight from the occulted portion of the stellar surface during transit. Briefly, we first removed the Keplerian motion and systemic velocity from the NEID CCF$_{\rm DI}$ (the cross-correlation function integrated over the entire stellar disk) using the best-fit parameters derived from the \exofasttwo\, analysis, and re-binned all CCF$_{\rm DI}$ into a common velocity grid in the stellar rest frame. Each CCF$_{\rm DI}$ was then normalized by its continuum level and scaled according to the flux predicted by the transit model. A master-out CCF$_{\rm DI}$ was created by averaging all out-of-transit CCFs. Finally, the local CCFs (CCF$_{\rm loc}$), representing the occulted stellar regions, were obtained by subtracting the scaled in-transit CCF$_{\rm DI}$ from the master-out CCF$_{\rm DI}$ (see Figure~\ref{fig:rrm}). Note that all CCF$_{\rm loc}$ with $\langle \mu \rangle < 0.4$ were excluded due to their weak signals, where $\mu = \cos\theta$ and $\theta$ is the center-to-limb angle.}

{The stellar velocities of the occulted starlight (hereafter local RVs) were derived by fitting Gaussian profiles to the CCF$_{\rm loc}$ using the \texttt{curve\_fit} routine implemented in \texttt{scipy} \citep{virtanen2020scipy}. The standard deviation of the CCF continuum was adopted as the uncertainty for each CCF data point, and the uncertainties of the fitted parameters were estimated from the covariance matrix returned by \texttt{curve\_fit}. The resulting local RVs are shown in Figure~\ref{fig:rrm}.}

{
We then fitted the local RVs using the \texttt{PyORBIT\footnote{\url{https://github.com/LucaMalavolta/PyORBIT}}} \citep{Malavolta2016,Malavolta2018} RRM module, which implements the algorithm of \citet{Cegla2016ReloadedRM}. Posterior sampling was performed using the dynamic nested sampling package \texttt{dynesty} \citep{Speagle2020}. During the fit, Gaussian priors on $P$, $T_c$, $a/R_\star$, and $i$ were adopted from the \exofasttwo\, results, while a Gaussian prior on $v\sin i_\star$ was taken from the \texttt{iSpec} analysis. Uniform priors $\mathcal{U}(0, 1)$ were used for the transformed limb-darkening coefficients $q_1$ and $q_2$, $\mathcal{U}(0, 1)$ for the differential rotation factor $\alpha_{\rm rot}$, $\mathcal{U}(-10, 10)$ for the center-to-limb convective variations (CLV) coefficients ($c_1$ and $c_2$), and $\mathcal{U}(-10, -1)$ (in log space) for the RV jitter term. During the sampling, we adopted $n_{\mathrm{live}} = 750$ live points and set \texttt{dlogz} = 0.01 as the stopping criterion, ensuring convergence once the change in the estimated Bayesian evidence ($\log Z$) became negligible.}

{We considered 12 model combinations accounting for the presence or absence of differential rotation, the inclusion of center-to-limb variation (linear or quadratic), and whether the stellar inclination was oriented towards ($i_\star < 90^\circ$) or away from ($i_\star > 90^\circ$) the observer. The modeling results are summarized in Table~\ref{tab:rrm_models}. The SB+CLV2 and DR+CLV2 (towards) models yield similarly lowest $\chi^2$ values, with a $\Delta$BIC of 4.6 ($<6$), indicating no strong statistical preference for one model over the other. Although the latter model has more degrees of freedom and a slightly higher BIC, its derived parameters ($\lambda$, $i_\star$, and $\psi$) are consistent with our classical RM analysis. Therefore, we adopt it as the preferred model, which gives $\alpha_{\rm rot} = 0.19^{+0.16}_{-0.13}$.
}

\begin{deluxetable*}{llcccccccccc}
\tabletypesize{\scriptsize}
\tablecaption{{Model comparison from RRM analysis.}\label{tab:rrm_models}}
\tablehead{
\colhead{Model} &
\colhead{$\lambda$} &
\colhead{$v\sin i_\star$} &
\colhead{$i_\star$} &
\colhead{$\psi$} &
\colhead{$\alpha_{\rm rot}$} &
\colhead{$c_1$} &
\colhead{$c_2$} &
\colhead{$\log \sigma_{\rm jitter}$} &
\colhead{BIC} &
\colhead{$\chi^2$} &
\colhead{$k$}  \\
\colhead{} &
\colhead{(deg)} &
\colhead{(km\,s$^{-1}$)} &
\colhead{(deg)} &
\colhead{(deg)} &
\colhead{} &
\colhead{(\kms)} &
\colhead{(\kms)} &
\colhead{} &
\colhead{} &
\colhead{}
}
\startdata
SB & $-79.2\pm1.8$ & $12.1^{+2.4}_{-1.7}$ & -- & -- & -- & -- & -- & $-3.2^{+2.2}_{-4.6}$ & 64.7 & 66.2 & 10 \\
SB CLV1 & $-79.8^{+2.1}_{-1.7}$ & $12.6^{+2.8}_{-2.0}$ & -- & -- & -- & $0.045^{+0.068}_{-0.033}$ & -- & $-2.0^{+1.0}_{-5.4}$ & 68.5 & 68.0 & 11 \\
SB CLV2 & $-75.8^{+2.7}_{-2.6}$ & $12.1^{+2.5}_{-1.9}$ & -- & -- & -- & $-3.5\pm3.2$ & $0.14^{+2.2}_{-2.3}$ & $-5.6^{+3.2}_{-3.1}$ & 30.7 & 15.5 & 12 \\
DR & $-74.7\pm2.9$ & $14.7^{+2.5}_{-2.0}$ & $77.4^{+8.8}_{-7.3}$ & $75.0^{+2.7}_{-2.6}$ & $0.65^{+0.082}_{-0.083}$ & -- & -- & $-5.6^{+3.3}_{-3.0}$ & 33.0 & 18.6 & 12 \\
DR (towards) & $-74.8^{+2.8}_{-2.5}$ & $14.9^{+2.5}_{-2.2}$ & $76.3^{+7.5}_{-7.2}$ & $75.1^{+2.3}_{-2.6}$ & $0.64^{+0.085}_{-0.069}$ & -- & -- & $-5.5^{+3.4}_{-3.1}$ & 32.6 & 18.3 & 12 \\
DR (away) & $-78.9^{+2.0}_{-2.2}$ & $14.5^{+2.2}_{-1.9}$ & $94.7^{+7.9}_{-3.5}$ & $79.0^{+2.3}_{-2.0}$ & $0.64^{+0.10}_{-0.086}$ & -- & -- & $-5.1^{+3.1}_{-3.3}$ & 36.2 & 24.1 & 12 \\
DR CLV1 & $-74.4^{+3.0}_{-2.9}$ & $14.6^{+2.7}_{-1.9}$ & $76.8^{+9.6}_{-6.7}$ & $74.7^{+2.8}_{-2.7}$ & $0.71^{+0.12}_{-0.089}$ & $0.40^{+0.65}_{-0.30}$ & -- & $-4.9^{+3.0}_{-3.5}$ & 35.5 & 21.5 & 13 \\
DR CLV1 (towards) & $-74.0^{+2.7}_{-2.3}$ & $14.9^{+2.2}_{-1.9}$ & $75.9^{+6.7}_{-6.0}$ & $74.4^{+2.2}_{-2.5}$ & $0.71^{+0.10}_{-0.095}$ & $0.42^{+0.68}_{-0.32}$ & -- & $-5.8^{+3.0}_{-2.7}$ & 35.1 & 19.2 & 13 \\
DR CLV1 (away) & $-78.7^{+2.3}_{-5.2}$ & $14.6^{+2.8}_{-2.0}$ & $94.5^{+12.3}_{-3.5}$ & $78.8^{+5.6}_{-2.3}$ & $0.74^{+0.21}_{-0.12}$ & $0.43^{+0.99}_{-0.33}$ & -- & $-5.7^{+3.2}_{-3.0}$ & 39.1 & 26.9 & 13 \\
DR CLV2 & $-75.7^{+6.1}_{-2.8}$ & $12.0^{+2.5}_{-2.4}$ & $126.4^{+31.1}_{-84.1}$ & $81.1^{+5.2}_{-6.9}$ & $0.28^{+0.69}_{-0.20}$ & $-1.7^{+4.4}_{-4.6}$ & $-2.0^{+3.4}_{-4.1}$ & $-5.0^{+2.9}_{-3.2}$ & 35.2 & 24.8 & 14 \\
\rowcolor[HTML]{D5E8D4} 
DR CLV2 (towards) & $-74.9^{+2.9}_{-2.7}$ & $12.3^{+2.6}_{-1.9}$ & 
$49.6^{+26.4}_{-34.1}$ & $78.8^{+6.8}_{-4.2}$ & $0.19^{+0.16}_{-0.13}$ &
$-4.5^{+4.0}_{-3.1}$ & $0.8^{+2.1}_{-2.7}$ & $-5.5^{+3.2}_{-2.9}$ & 
{35.3} & {15.6} & 14 \\
DR CLV2 (away) & $-76.5^{+6.4}_{-3.3}$ & $12.1^{+2.5}_{-2.2}$ & $141.7^{+16.5}_{-20.5}$ & $82.5^{+3.8}_{-6.1}$ & $0.40^{+0.71}_{-0.26}$ & $-0.21^{+4.2}_{-4.5}$ & $-3.5^{+3.4}_{-3.7}$ & $-5.4^{+3.1}_{-2.9}$ & 35.3 & 17.7 & 14 \\
\enddata
\tablenotetext{}{\textbf{Notes.} 
SB = solid-body without differential rotation; DR = differential rotation; CLV = center-to-limb variation term.
The ``towards'' and ``away'' labels denote models where the stellar inclination $i_\star$
is constrained towards $i_\star < 90\degrees$ or away $i_\star > 90\degrees$ from the sky-plane. The favored solution has been highlighted.
}
\end{deluxetable*}
\onecolumngrid

\section{ \teff\,--$\lambda$ for near-polar sub-Saturns from secular resonance crossing}\label{appendix:simulation}

\citealt{Petrovich2020} demonstrated that a secular inclination resonance can arise in multi-planet systems embedded in a dissipating protoplanetary disk. In this scenario, the outer planet remains gravitationally coupled to the disk, whose mass and gravitational potential decrease over time, causing the outer planet’s nodal precession rate to evolve. Simultaneously, the inner planet experiences nodal precession primarily driven by the star’s rotationally induced quadrupole moment. As the disk dissipates, a secular resonance is crossed, enabling efficient angular momentum exchange between the planets. Due to conservation of the angular momentum deficit (AMD, \citealt{Laskar1997}), the resonance transfers inclination predominantly to the inner planet, which typically has lower orbital angular momentum, allowing it to acquire significant stellar obliquity while the outer planet and disk remain nearly coplanar.

Deriving stellar‐obliquity predictions from secular‐resonance crossing requires accurate knowledge of stellar parameters, including radius, mass, \teff\, and, most critically, the rotation period. We adopted rotation periods from \cite{McQuillan2014} and took the remaining parameters from the Kepler Stellar Table in the NASA Exoplanet Archive. Because \cite{McQuillan2014} reports reliable rotation periods only for stars cooler than 7000 K, we retained the catalog values of radius, mass, and \teff\ for targets with \teff\,$>$ 7000 K and, to be conservative, assumed their rotation periods to be uniformly distributed between 0.5 and 1 day (e.g., \citealt{Balona2021}). For the Love number ($k_{2}$), we adopted representative values from the stellar models constructed by \cite{Claret2023} for stars with effective temperatures between 4000 and 8000 K. Using these values we then modeled the Love number as a decaying power law: $k_2\approx0.075(T_{\rm eff}/4000)^{-4}$. The procedure results in 40,000 groups of stellar parameters, ranging from 4000 to 8000 K, for the analytic calculation of the stellar obliquity prediction from \citet{Petrovich2020} given by:
\begin{equation}
 \psi_{\rm final} =
\begin{cases}
 \arcsin\sqrt{(4+4\eta_{\star}+\eta_{\rm GR})/(10+5\eta_{\star})}, & \text{if } \eta_{\rm GR}<6+\eta_\star, \\
  \pi/2,  & \text{if } \eta_{\rm GR}\geq 6+\eta_\star
\end{cases}
\label{eq:inc_crit}
\end{equation}
where
\begin{equation}
    \eta_{\star}=\frac{2J_2R_{\star}^2M_{\star}a_{\rm out}^3(1-e_{\rm out})^{3/2}}{M_{\rm out}a_{\rm in}^5},
    \label{eq:eta_star}
\end{equation}
corresponds to the relative strength of the stellar quadrupole parametrized by $J_2=k_2\Omega^2_\star\ R^3_\star/(3GM_\star)$ with respect to the two-planet interactions, and
\begin{equation}
    \eta_{\rm GR}=\frac{8GM_{\star}^2a_{\rm out}^3(1-e_{\rm out})^{3/2}}{c^2M_{\rm out}a_{\rm in}^4},
    \label{ec:eta_GR}
\end{equation}
represents the relative strength of the general relativity corrections with respect to the two-planet interactions. From equation \ref{eq:inc_crit}, one can easily check that when the stellar quadrupole dominates ($\eta_{\star}\gg1$) one obtains a critical inclination of $\sin^{-1}(4/5)^{1/2}\approx63.4^{\circ}$.

To incorporate the effects of measurement uncertainties, we performed a Monte Carlo resampling procedure based on binned stellar parameters. Specifically, we first partitioned the \teff\, data into 5 K-wide bins. Within each bin, we computed the mean and standard deviation of each stellar parameter of interest (e.g., stellar mass, radius, and rotation period). Furthermore, to mitigate the influence of outliers, we applied a 1-$\sigma$ clipping filter to the residuals relative to a non-parametric smoothed curve computed with the \texttt{lowess} function from the \texttt{statsmodels} package \citep{seabold2010statsmodels}.

In the simulation, we adopted a representative configuration with an inner sub-Saturn initially at the equatorial plane with a semi-major axis of 0.05 AU and a planetary mass of 0.2 $M_{\rm J}$. For the outer companion, we chose a characteristic mass of 1 $M_{\rm J}$, a semi-major axis of 2 AU, which relates to the semi-major axis of the inner planet according to
\begin{equation}
    a_{\rm out}\gtrsim2\left(\frac{a_{\rm in}}{0.05}\right)^{4/3}\text{ }\rm{AU},
\end{equation}
and a mutual inclination of 6.51\degrees. This outer companion ensures that $\eta_{\star}\gg1$ for the rotation rates at large $T_{\rm eff}$, allowing a significant decrease of $\psi$ past the Kraft break (below the break $a_{\rm out}\gtrsim2$ AU would suffice to get a polar obliquity for the inner planet due to relativistic precession).

Notice that, in the simulation, we assumed a stellar inclination of 90\degrees, so the resulting $\psi$ equals $\lambda$. Using the synthetic dataset and this configuration, we re-evaluated the correlation between \teff\, and stellar obliquity for polar-orbiting sub-Saturns, which is shown in the Figure~\ref{fig:teff_lam}. Our simulation indicates that secular resonance crossing can lead to strictly polar-orbiting sub-Saturns in cool-star systems. However, due to the rapid rotation of hot stars, the resulting stellar quadrupole can suppress the general relativistic apsidal precession, causing sub-Saturns to linger around 65\degrees.

\textbf{}
\bibliography{main}{}
\bibliographystyle{aasjournal}

\end{document}